\documentclass[twocolumn,showpacs,amsmath,amssymb,prl]{revtex4}
\usepackage{graphicx}
\usepackage{dcolumn}
\usepackage{bm}
\newcommand{\VECr}{{\boldsymbol{r}}}
\begin{document}
\title{Three-Particle Correlations in Simple Liquids}
\author{K. Zahn, G. Maret, C. Ru\ss{} and H.H. von Gr\"unberg}
\affiliation{Universit\"at Konstanz, Fachbereich Physik, P.O.B. 5560,
  78457 Konstanz, Germany}
\date{\today}
\begin{abstract}
  We use video microscopy to follow the phase-space trajectory of a
  two-dimensional colloidal model liquid and calculate three-point
  correlation functions from the measured particle configurations.
  Approaching the fluid-solid transition by increasing the strength of
  the pair-interaction potential, one observes the gradual formation
  of a crystal-like local order due to triplet correlations, while
  being still deep inside the fluid phase. Furthermore, we show that in
  a strongly interacting system the Born-Green equation can be
  satisfied only with the full triplet correlation function but not
  with three-body distribution functions obtained from superposing
  pair-correlations (Kirkwood superposition approximation).
\end{abstract}
\maketitle

Our current understanding of the structure of simple fluids is based
on the $n$-body distribution functions $g^{(n)}$, measuring the
probability density of finding two, three, and more particles at
specified positions in space. When the total potential energy of a
liquid is given by a sum of pair-potentials, all of its thermodynamic
properties can be calculated by means of the pair-correlation function
$g(r)\equiv g^{(2)}(r)$ and its density ($\rho$) and temperature ($T$)
derivatives.  However, the latter two quantities, $\partial g(r)/
\partial \rho$ and $\partial g(r)/ \partial T$, explicitly depend on
the triplet correlation function, even if the particle interactions
are only pair-wise additive \cite{schofield66}. Explicit knowledge of
triplet correlations is also required in perturbation theories for
static fluid properties \cite{stell74}, in theories of transport
properties \cite{scherwinski90}, of solvent reorganization processes
around solutes \cite{lazaridis00}, of systems under shear-flow
\cite{dhont7599}, but also to understand the structural properties of
a 2D amorphous system \cite{koenig}. Most of our knowledge on triplet
correlations come from computer simulation studies of
hard-sphere-fluids \cite{alder64}, Lennard-Jones fluids
\cite{rahman64,mcneil83} and electrolyte systems
\cite{linse1991}. In the
overwhelming majority, these papers are concerned with testing
Kirkwood's superposition approximation (KSA) \cite{kirkwood35} for the
triplet distribution function.  By contrast, semi-analytical theories
for $g^{(3)}$ beyond the KSA are rather rare \cite{mcneil83,rice65}.
However, despite the long-standing theoretical interest in its
properties, it has never been possible to measure three-particle
correlations directly.  Indirect ways to identify higher-order
correlations in scattering data have been suggested for instance in
\cite{dietrich}. An alternative, but also indirect way to obtain
experimental information on $g^{(3)}$ is based on the relationship
between the isothermal pressure derivative of the fluid structure
factor $\partial S(q)/\partial P$ and the triplet distribution
function \cite{schofield66}, a relationship which has been
systematically exploited by Egelstaff and co-workers in rare-gas
systems \cite{egel73}. The present Letter reports on the first direct
measurement of $g^{(3)}$ in a two-dimensional colloidal model liquid
with well-defined pair-interaction potentials.

\begin{figure}
\includegraphics[width=0.45 \textwidth]{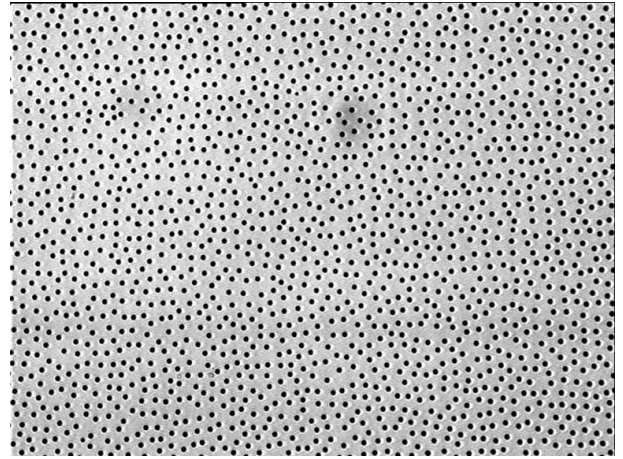}
\caption{\label{fig0} 
  A typical image ($500 \times 380\: \mu m$) of our two-dimensional
  colloidal model system with paramagnetic colloids of $d=4.7 \mu m$
  diameter. The particles interact via a potential $\sim \Gamma/r^{3}$
  in which the interaction strength $\Gamma$ can be conveniently varied
  through the external magnetic field.}
\end{figure}

The preparation of the samples and the experiments have been carried
out as described in \cite{zahn99}: Spherical colloids (diameter $d=4.7
\:\mu m$) are confined by gravity to a water/air interface whose
flatness can be controlled within less than a micron. The field of
view has a size of $520 \times 440 \: \mu m$ containing typically
about $10^{3}$ particles. The particles are super-paramagnetic due to
Fe$_{2}$O$_{3}$ doping. A magnetic field $B$ applied perpendicular to
the air/water interface induces in each particle a magnetic moment $M=
\chi B$ which leads to a repulsive dipole-dipole pair-interaction
energy of $\beta u(r) = \Gamma/ (\sqrt{\pi\rho}r)^{3}$ with the
interaction strength given by $\Gamma = \beta (\mu_{0}/4 \pi) (\chi
B)^{2} (\pi \rho)^{3/2}$ ($\beta = 1/kT$ the inverse temperature,
$\chi$ the susceptibility). This is the only relevant contribution to
the interparticle-potential which is hence conveniently and reversibly
adjustable by the external field $B$. A typical 
snapshot of our system is given in Fig.~(\ref{fig0}).
$\Gamma$ is the only parameter determining the phase-behavior of the
system: for $\Gamma < 57$ the system is liquid, for $\Gamma > 60$ it
is solid, and in between, i.e. for $57 < \Gamma < 60$, it shows a
hexatic phase \cite{zahn99}.  We here analyze three different
$\Gamma$($\Gamma = 4,14,46$), where the system is deep in the liquid
phase, and use for each (well equilibrated) system about 200
statistically independent configurations with approximately 500
particles, recorded using digital video-microscopy with subsequent
image-processing on the computer. From the measured particle
configurations, $g^{(3)}$ is obtained by computing the average count
per configuration of a particular kind of triplet, divided then by the
appropriate normalizing factor. Details of this calculation will be
given elsewhere \cite{russ3}.

Triplet correlations can be characterized by the ratio between the
full triplet distribution function $g^{(3)}$ and its approximated form
based on the KSA $g^{(3)}_{SA} =
g(\VECr_{1})g(\VECr_{2})g(\VECr_{3})$. This ratio is given by what is
called the triplet correlation function, denoted here by
$G(\VECr_{1},\VECr_{2},\VECr_{3})$. Thus, $g^{(3)}= g_{SA}^{(3)}G$.
All pair correlations in $g^{(3)}$ are included in $g^{(3)}_{SA}$,
while the extent of the intrinsic correlations due to the simultaneous
presence of a triplet of particles at positions $\VECr_{1}$,
$\VECr_{2}$, $\VECr_{3}$ is quantified through the function $G$, which
thus defines the local structure of the fluid beyond that expressed by
the pair-correlation functions. Introducing $\beta w^{(m)}= - \ln
g^{(m)}$, $g^{(3)}= g_{SA}^{(3)}G$ transforms into
\begin{equation}
  \label{eq:2}
w^{(3)}(\VECr_{1},\VECr_{2},\VECr_{3}) =
w^{(2)}(\VECr_{1})+w^{(2)}(\VECr_{2})+w^{(2)}(\VECr_{3}) - \ln G /\beta
\end{equation}
with $w^{(2)}$ and $w^{(3)}$ being the two- and three-particle
potential of mean force, respectively.  The equation shows that $- \ln
G \equiv \beta \Delta w^{(3)}$ plays the role of a three-body
potential, measuring the (extra correlation) energy of three
correlated particles relative to the energy of superposed correlated
pairs of particles.
\begin{figure}
\includegraphics[width=0.45 \textwidth]{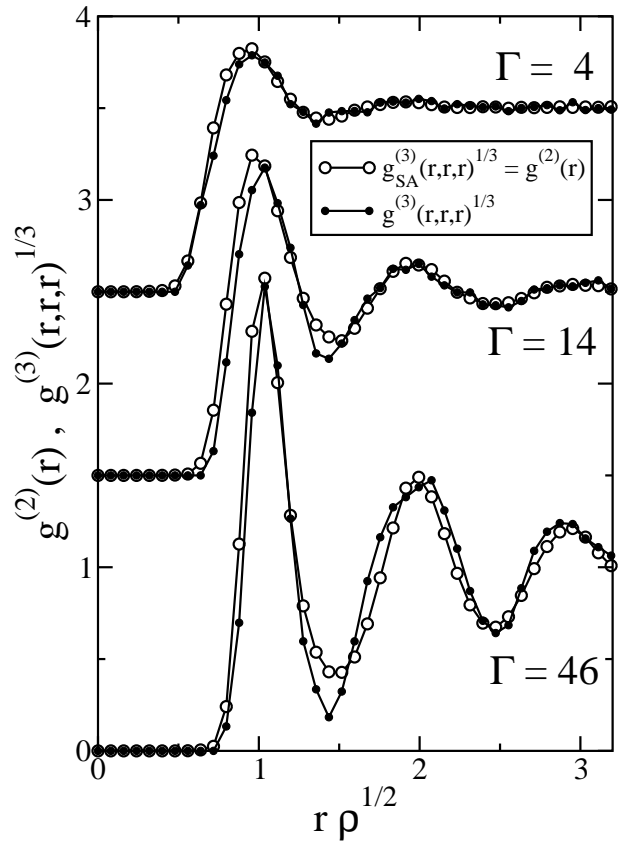}
\caption{\label{fig1} 
  Triplet distribution functions $g^{(3)}$ as a function of the side
  length of an equilateral triangle, as computed from measured
  particle configurations for different $\Gamma$. $g_{SA}^{(3)}$ is
  the triplet distribution function in the Kirkwood superposition
  approximation which on taking the cubic root,
  $\sqrt[3]{g_{SA}^{(3)}(r,r,r)}$, becomes the radial distribution
  function $g(r)$.}
\end{figure}

\begin{figure}
\includegraphics[width=0.45 \textwidth]{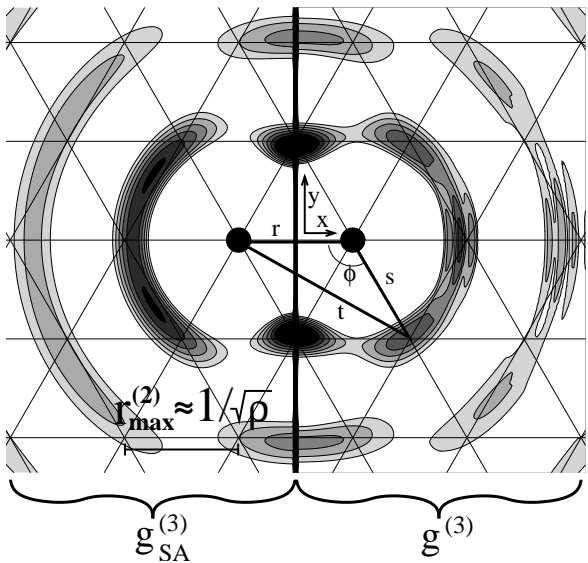}
\caption{\label{fig2} 
  Distribution functions $g^{(3)}(r=r^{(2)}_{max},s(x,y),t(x,y))$
  (right half of the figure) and
  $g_{SA}^{(3)}(r=r^{(2)}_{max},s(x,y),t(x,y))$ (left half of the
  figure) in the $(x,y)$-plane ($\Gamma = 46$). The missing half of
  each distribution is just the mirror-image of the one actually
  plotted.  The constant $g(r^{(2)}_{max})$ is subtracted from the
  distributions and only positive values are plotted with a grey-level
  scheme between white (zero) and black (max. value).}
\end{figure}

\begin{figure}
\includegraphics[width=0.45 \textwidth]{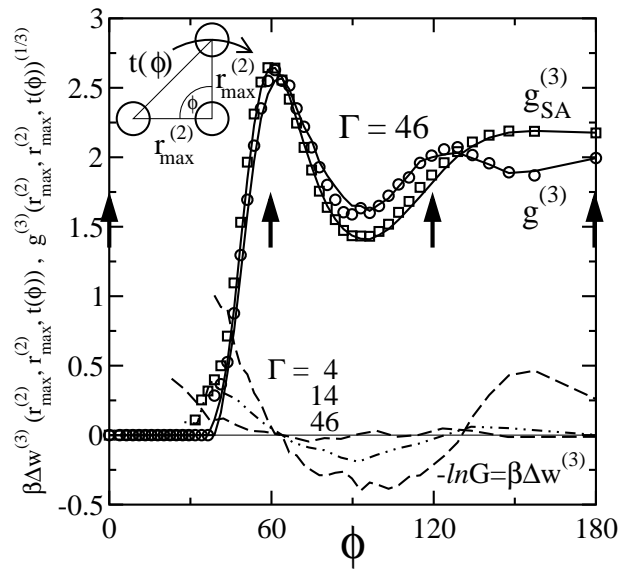}
\caption{\label{fig3} 
  $g^{(3)}$ and $g_{SA}^{(3)}$ from Fig.~(\ref{fig2}) for fixed values
  of $r=r^{(2)}_{max}$ and $s=r^{(2)}_{max}$, as a function of the
  angle $\phi$ (see inset). Symbols (solid lines) for
  distributions generated from measured (MC-simulated) configurations.
  Also given is the logarithm of the triplet correlation function $G$,
  which is related to the triplet correlation energy $\Delta w^{(3)}$,
  for $\Gamma=4$ (dashed line), $\Gamma=14$ (dashed-dotted line) and
  $\Gamma=46$ (dashed line).}
\end{figure}

In an homogeneous, isotropic system, $g^{(3)}$ depends on only three
independent variables, chosen here to be $r=|\VECr_{1} - \VECr_{2}|$,
$s=|\VECr_{2} - \VECr_{3}|$ and $t=|\VECr_{1} - \VECr_{3}|$.
Fig.~(\ref{fig1}) shows three-particle distribution functions in the
equilateral triangle geometry for all three $\Gamma$'s considered. To
allow comparison with the radial distribution function $g(r)$ we have
taken the cubic root $\sqrt[3]{g^{(3)}(r,r,r)}$ so that
$\sqrt[3]{g_{SA}^{(3)}(r,r,r)} = g(r)$. It is evident that the KSA,
while working satisfactorily at low $\Gamma$, fails to reproduce the
fine-structure of the triplet distribution function at higher values
of $\Gamma$. Obviously, correlations beyond the level of
pair-correlations become important at higher $\Gamma$. To visualize
$g^{(3)}$ in two dimensions, $r$ is fixed in the following to the
distance $r^{(2)}_{max}$ where $g(r)$ has its first peak ($\approx
1/\sqrt{\rho}$), so that $g^{(3)}(r^{(2)}_{max},s,t)$ varies just with
$s=s(x,y)$ and $t=t(x,y)$ and can thus be plotted in the $(x,y)$-plane
in form of a contour-plot. This is done in Fig.~(\ref{fig2}) for the
$\Gamma = 46$ measurement.  We show in the left half of the figure
$g^{(3)}_{SA}$ and contrast it to the full three-particle distribution
function $g^{(3)}$, plotted in the right half of the figure. $g^{(3)}$
approaches $g(r^{(2)}_{max})$ for large values of $x^{2}+y^{2}$. To
keep the figure as clear as possible, we plotted just those parts of
$g^{(3)} - g(r^{(2)}_{max})$ and $g^{(3)}_{SA} - g(r^{(2)}_{max})$
that are larger than zero. The stripes that can be seen especially
close to the x-axis result from the transformation
$g^{(3)}(r_{max},s,t)$ to $g^{(3)}(r_{max},x,y)$ and appear due to
limited statistics. A hexagonal lattice with a lattice constant $a=
r^{(2)}_{max}$ is superposed.
Both distributions $g^{(3)}$ and $g_{SA}^{(3)}$ reveal that the
neighbors of the two central particles have positions which show a
certain correspondence to the crystalline lattice points. However,
while the banana-like structure of $g^{(3)}_{SA}$ reflects just the
coordination shells of the lattice, the full distribution function
$g^{(3)}$ shows a well-developed, angular dependent substructure, with
individual peaks for every lattice point in the first coordination shell.
Fig.~(\ref{fig3}) shows $g^{(3)}$ and $g_{SA}^{(3)}$ of
Fig.~(\ref{fig2}) along the line
$(r=r_{max}^{(2)},s=r_{max}^{(2)},t=t(\phi))$, which is a circle of
radius $r_{max}^{(2)}$ around the right particle in Fig.~(\ref{fig2}),
passing through all lattice points of the particle's first
coordination shell (arrows in Fig.~(\ref{fig3}) mark positions of
lattice points). It can be clearly seen that $g^{(3)}$ develops peaks
at the lattice points while $g_{SA}^{(3)}$ completely fails to reflect
the hexagonal structure. Also given is the function $- \ln G$, i.e.
$\Delta w^{(3)}$, of eq.~(\ref{eq:2}), now for all three values of
$\Gamma$ studied here. It is evident how $\Delta w^{(3)}$ gradually
forms on increasing $\Gamma$, with values up to one $kT$ (in other
regions of the $(r,s,t)$-space we find energies as high as $4$ $kT$!).
It is also seen that the regions of attractive and repulsive
correlation energies $\Delta w^{(3)}$ correspond to the correcting
effect which the function $G$ has on $g_{SA}^{(3)}$ to ensure that
$g^{(3)}$ adapts locally to the hexagonal symmetry.  
We conclude that it is an effect entirely due to three-particle
correlations, i.e. due to the function $G$, which is responsible for
the observed formation of a crystal-like local environment around
particles well below the freezing transition. We also performed
Monte-Carlo (MC) simulations using the above-given pair-potential
$\beta u(r) = \Gamma/ (\sqrt{\pi\rho}r)^{3}$ (with a better
statistic than in the experiment: 500 configurations with 2000
particles, periodic boundary conditions). The almost perfect agreement
between the distribution functions based on the MC-data (solid lines
in Fig.~(\ref{fig3})) and on the experimental configurations (symbols
in Fig.~(\ref{fig3})), demonstrates that our model liquid
consists of particles interacting solely via pair-wise additive and
precisely known potentials.

Furthermore, we carried out MC simulations for the solid phase
($\Gamma = 80$), starting from a perfect hexagonal lattice, and
compared the resulting triplet distribution function with the
experimental one for the $\Gamma = 46$ measurement in the liquid
phase, see \cite{russ3}. The distributions look quite similar: as
regards the correlations between the central pair and the first
coordination shell (in a plot like that in Fig.~(\ref{fig2})), there
is hardly any difference between the liquid and the solid phase.
Pronounced differences are observable, however, in the second shell:
in the liquid phase the next nearest neighbors are broadly distributed
midway between adjacent lattice nodes (see Fig.~(\ref{fig2})), while
the $\Gamma = 80$ distribution correlates much better with the lattice
structure. However, even for $\Gamma = 80$ this correspondence is far
from perfect; it is well developed along the $y$-direction, but becomes
worse on increasing $\phi$ to $180^\circ$ where there is still an
extended smeared-out distribution showing no clear preference for
certain lattice points. Clearly, approaching $T \to 0$ ($\Gamma \to
\infty$), one will ultimately observe peaks in $g^{(3)}$ positioned
exclusively on the lattice points. We should remark that in three dimensions a
similar correspondence between the peaks in $g^{(3)}$ and an
underlying crystal lattice should be much harder to find. In 3D, every
triplet of particle lies, of course, also in a plane, and can
accordingly be plotted as in Fig.~(\ref{fig2}).  However, then there
is not one, but a superposition of many possible lattice planes that
one has to compare this distribution with.

\begin{figure}
\includegraphics[width=0.45 \textwidth]{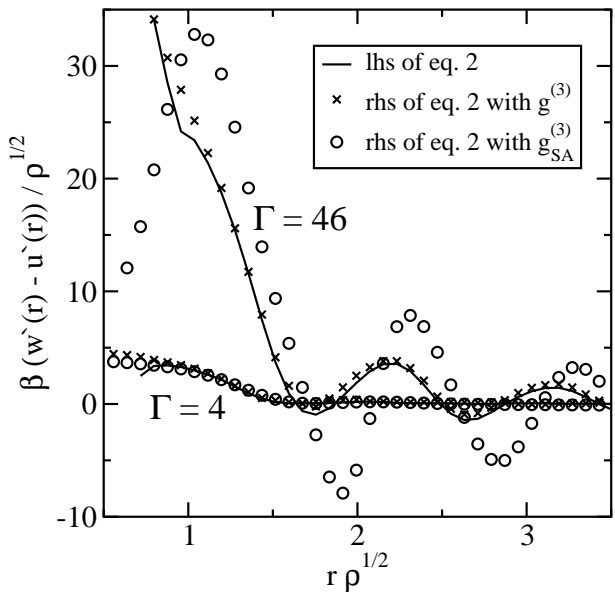}
\caption{\label{fig4} 
Test of Kirkwood's approximation using experimentally
determined three-particle distribution functions ($\Gamma=46$ and
$\Gamma=4$). Solid lines for the left hand side of the Born-Green
equation (eq.~\ref{eq:3}), symbols for the right hand side, evaluated
using the full triplet distribution function $g^{(3)}$ (crosses)
and the distribution function $g_{SA}^{(3)}$ (open circles) which is based
on Kirkwood's superposition approximation.}
\end{figure}

To demonstrate that triplet correlations are significant not only
locally, but also when integrated over the whole volume we consider
the Born Green equation \cite{born46},
\begin{equation}
  \label{eq:3}
\frac{\partial w^{(2)}(r_{12})}{\partial \VECr_{1}} 
-  \frac{\partial u(r_{12})}{\partial \VECr_{1}}
= \rho \int \frac{\partial u(r_{13})}{\partial \VECr_{1}} 
\frac{g^{(3)}(\VECr_{1},\VECr_{2},\VECr_{3})}{g(r_{12})} d\VECr_{3}\:,
\end{equation}
relating the difference between the mean force and the direct
pair-force to an integral over the force on particle~1 due to a
third particle at $\VECr_{3}$, weighted by the probability $\rho g^{(3)}
d\VECr_{3}/g(r_{12})$ of finding this particle in $d\VECr_{3}$ at
$\VECr_{3}$ when it is known that other particles are located at
$\VECr_{1}$ and $\VECr_{2}$. This equation is exact if pairwise
interactions can be assumed. To illustrate the importance of
three-particle correlations, we numerically computed the right-hand
side of eq.~(\ref{eq:3}) using both the full and the approximated
triplet function, $g^{(3)}$ and $g_{SA}^{(3)}$, of the $\Gamma=4$ and
$\Gamma=46$ measurement and compared it in Fig.~(\ref{fig4}) to the
left-hand side of eq.~(\ref{eq:3}), evaluated using $u(r)$ and $g(r)$.
For the strongly interacting system ($\Gamma = 46$)
the KSA fails completely.  Three-particle correlations are
thus seen to be important not only to obtain locally the correct
structure, but also to obtain globally the correct difference between
mean and direct force via the Born-Green equation.

Acknowledgment: We are grateful to Gerd Haller for providing the photograph
of the sample in Fig.~(\ref{fig0}).

\end{document}